\documentclass[amsmath,amssymb,aps,letterpaper,prl,twocolumn,preprintnumbers]{revtex4}

\usepackage[final]{graphicx}
\makeatletter
\newlength{\apb@width}
\newcommand{\autoparbox}[2][c]{\settowidth{\apb@width}{#2}\parbox[#1]{\apb@width}{#2}}
\newcommand{\includegraphicsbox}[2][]{\autoparbox{\includegraphics[#1]{#2}}}
\makeatother

\allowdisplaybreaks[3]

\makeatletter \@addtoreset{paragraph}{section} \makeatother

\let\oldPhi=\Phi
\let\oldPsi=\Psi
\let\oldGamma=\Gamma
\let\oldDelta=\Delta
\let\oldSigma=\Sigma
\let\oldLambda=\Lambda
\let\oldTheta=\Theta
\let\oldPi=\Pi
\renewcommand{\Phi}{\mathnormal{\oldPhi}}
\renewcommand{\Psi}{\mathnormal{\oldPsi}}
\renewcommand{\Gamma}{\mathnormal{\oldGamma}}
\renewcommand{\Sigma}{\mathnormal{\oldSigma}}
\renewcommand{\Delta}{\mathnormal{\oldDelta}}
\renewcommand{\Theta}{\mathnormal{\oldTheta}}
\renewcommand{\Lambda}{\mathnormal{\oldLambda}}
\renewcommand{\Pi}{\mathnormal{\oldPi}}

\newcommand{\charge}{\mathcal{Q}}

\newcommand{\loc}{\mathcal{L}}
\newcommand{\boost}[1]{\mathcal{B}[#1]}
\newcommand{\boostLeft}[1]{\mathcal{B}_\Left[#1]}
\newcommand{\boostRight}[1]{\mathcal{B}_\Right[#1]}
\newcommand{\biloc}[2]{[#1|#2]}
\newcommand{\ueber}[2]{\langle#1|#2\rangle}

\newcommand{\superN}{\mathcal{N}}
\newcommand{\SR}{0}

\newcommand{\bulk}{\mathrm{bulk}}
\newcommand{\bound}{\mathrm{bdr}}

\newcommand{\Left}{{\text{\tiny L}}}
\newcommand{\Right}{{\text{\tiny R}}}
\newcommand{\Span}{{\text{\tiny L\&R}}}
\newcommand{\conn}{\mathcal{X}}
\newcommand{\id}{\mathcal{I}}

\newcommand{\rot}{\mathcal{G}}

\newcommand{\order}[1]{\mathcal{O}(#1)}

\newcommand{\specleg}{%
\unitlength2pt
\scalebox{.45}{
\begin{picture}(3,11)(1.4,.2)%
\thicklines
\put(2,0){\line(0,1){8}}%
\put(2.1,0){\circle*{2}}%
\put(2.1,8){\circle*{2}}%
\end{picture}%
}}

\newcommand{\twospecleg}{%
\unitlength2pt
\scalebox{.45}{
\begin{picture}(10,11)(0,.2)%
\thicklines
\put(2,0){\line(0,1){8}}%
\put(2.1,0){\circle*{2}}%
\put(2.1,8){\circle*{2}}%
\put(8,0){\line(0,1){8}}%
\put(8.1,0){\circle*{2}}%
\put(8.1,8){\circle*{2}}%
\end{picture}%
}
}

\newcommand{\permsym}{%
\unitlength2pt
\scalebox{.45}{
\begin{picture}(10,11)(0,.2)%
\thicklines
\put(2,0){\line(3,4){5.7}}%
\put(2.1,0){\circle*{2}}%
\put(2.1,8){\circle*{2}}%
\put(8,0){\line(-3,4){5.7}}%
\put(7.9,0){\circle*{2}}%
\put(7.9,8){\circle*{2}}%
\end{picture}%
}}

\newcommand{\threeonetwo}{%
\unitlength2pt
\scalebox{.45}{
\begin{picture}(12,11)(1.5,.2)%
\thicklines
\put(2,0){\line(2,3){5.5}}%
\put(2.1,0){\circle*{2}}%
\put(2.1,8){\circle*{2}}%
\put(7,0){\line(2,3){5.5}}%
\put(7.1,0){\circle*{2}}%
\put(7.1,8){\circle*{2}}%
\put(12,0){\line(-5,4){10}}%
\put(12,0){\circle*{2}}%
\put(12,8){\circle*{2}}%
\end{picture}%
}}

\newcommand{\twothreeone}{%
\unitlength2pt
\scalebox{.45}{
\begin{picture}(12,11)(1.5,.2)%
\thicklines
\put(2,0){\line(5,4){10}}%
\put(2.1,0){\circle*{2}}%
\put(2.1,8){\circle*{2}}%
\put(7,0){\line(-2,3){5.5}}%
\put(7,0){\circle*{2}}%
\put(7,8){\circle*{2}}%
\put(12,0){\line(-2,3){5.5}}%
\put(12,0){\circle*{2}}%
\put(12,8){\circle*{2}}%
\end{picture}%
}}

\newcommand{\onethreetwo}{%
\unitlength2pt
\scalebox{.45}{
\begin{picture}(12,11)(1.5,.2)%
\thicklines
\put(2,0){\line(0,0){8}}%
\put(2.1,0){\circle*{2}}%
\put(2.1,8){\circle*{2}}%
\put(7,0){\line(2,3){5.5}}%
\put(7,0){\circle*{2}}%
\put(7,8){\circle*{2}}%
\put(12,0){\line(-2,3){5.5}}%
\put(12,0){\circle*{2}}%
\put(12,8){\circle*{2}}%
\end{picture}%
}}

\newcommand{\twoonethree}{%
\unitlength2pt
\scalebox{.45}{
\begin{picture}(12,11)(1.5,.2)%
\thicklines
\put(2,0){\line(2,3){5.5}}%
\put(2.1,0){\circle*{2}}%
\put(2.1,8){\circle*{2}}%
\put(7,0){\line(-2,3){5.5}}%
\put(7,0){\circle*{2}}%
\put(7,8){\circle*{2}}%
\put(12,0){\line(0,1){8}}%
\put(12,0){\circle*{2}}%
\put(12,8){\circle*{2}}%
\end{picture}%
}}

\ifx\genfrac\sdflkaj

\else

\fi
\newcommand{\sfrac}[2]{{\textstyle\frac{#1}{#2}}}
\newcommand{\half}{\sfrac{1}{2}}
\newcommand{\ihalf}{\sfrac{i}{2}}

\newcommand{\bigbrk}[1]{\bigl(#1\bigr)}
\newcommand{\brk}[1]{(#1)}

\newcommand{\bigcomm}[2]{\big[#1,#2\big]}
\newcommand{\comm}[2]{[#1,#2]}

\newcommand{\range}[1]{|#1|}

\newcommand{\ket}[1]{|#1\rangle}

\newcommand{\alg}[1]{\mathfrak{#1}}

\newcommand{\beq}{\begin{equation}}
\newcommand{\eeq}{\end{equation}}

\def\[{\begin{equation}}
\def\]{\end{equation}}
\def\<{\begin{eqnarray}}
\def\>{\end{eqnarray}}

\makeatletter
\def\mr@ignsp#1 {\ifx\:#1\@empty\else #1\expandafter\mr@ignsp\fi}%
\newcommand{\multiref}[1]{\begingroup
\xdef\mr@no@sparg{\expandafter\mr@ignsp#1 \: }%
\def\mr@comma{}%
\@for\mr@refs:=\mr@no@sparg\do{\mr@comma\def\mr@comma{,}\ref{\mr@refs}}%
\endgroup}
\makeatother

\newcommand{\hypref}[2]{\ifx\href\asklfhas #2\else\href{#1}{#2}\fi}

\newcommand{\Tabref}[1]{Table~\multiref{#1}}

\newcommand{\Figref}[1]{Figure~\multiref{#1}}

\ifx\href\asklfhas\newcommand{\href}[2]{#2}\fi

\usepackage{color}
\begin{document}


\preprint{LPT-ENS-12-01}
\preprint{AEI-2012-000}

\title{Recursion Relations for Long-Range Integrable Spin Chains\\ with Open Boundary Conditions}

\author{Florian Loebbert}
 \email{loebbert@nbi.dk}
\affiliation{
Niels Bohr International Academy,
The Niels Bohr Institute,
 Blegdamsvej 17, 2100 Copenhagen, Denmark
}
\affiliation{%
Laboratoire de Physique Th\'eorique,
\'Ecole Normale Sup\'erieure,
24 rue Lhomond, 75005 Paris, France
}
\affiliation{%
Max-Planck-Institut f\"ur Gravitationsphysik,
Albert-Einstein-Institut,
Am M\"uhlenberg 1, 14476 Potsdam, Germany
}

\date{\today}

\begin{abstract}
It is well known that integrable charges for short-range (e.g.\ nearest-neighbor) spin chains with periodic boundary conditions can be recursively generated by a so-called boost operator. In the past, this iterative construction has been generalized to periodic long-range spin chains as they appear in the context of the gauge/gravity correspondence. Here we introduce recursion relations for open long-range spin chain charges converting a short-range into a long-range integrable model.
\end{abstract}

\maketitle


\section{Introduction}

Integrable spin chains provide a universal tool for a variety of different physical problems. They are a natural concept in condensed matter physics and of great importance for the most prominent examples of the gauge/gravity duality, where a long-range generalization of the Heisenberg spin chain led to impressive progress \cite{Beisert:2010jr}. The interaction range of the commuting charges of this integrable chain increases with the perturbative order of the quantum field theory coupling constant which gives a beautiful example of a fruitful marriage between different areas of physics.

While the integrable spin chains employed for the spectral problem of the gauge/gravity duality typically have periodic boundaries, open boundary conditions arise in different important contexts: They describe so-called giant graviton states \cite{Berenstein:2005vf,Hofman:2007xp}, addition of fundamental matter to superconformal theories \cite{Erler:2005nr} or operator insertions into Wilson loops \cite{Drukker:2006xg}. In each case their discovery tremendously simplifies the solution of the underlying problem. 

\begin{figure}[t]
 \Large $\sum\limits_{a}$\quad\includegraphicsbox[scale=.8]{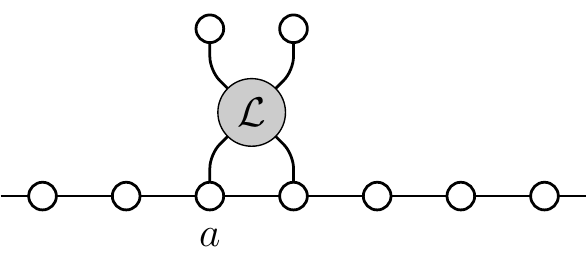}
\caption{Local operator of range 2 acting on a spin chain.}
\label{fig:locop}
\end{figure}

For most \emph{periodic} short-range spin chains the characteristic integrable charges may be obtained from the Hamiltonian $\charge_2$ via a master symmetry of the form \cite{Tetelman:1981xx,Sogo:1983aa}
\begin{equation}\label{eq:srboost}
\charge_{r+1}=\sfrac{i}{r}\,\comm{\boost{\charge_2}}{\charge_r},\qquad r=2,3,\dots.
\end{equation}
The so-called boost operator $\boost{\charge_2}$ is discussed below. While the short-range recursion \eqref{eq:srboost} for periodic chains was generalized to long-range chains in \cite{Bargheer:2008jt,Bargheer:2009xy}, neither short- nor long-range recursions are known for \emph{open} spin chains where only the even half $\charge_{2r}$ of the charges is conserved, cf.\ \cite{Grabowski:1996qf}. In this work we introduce recursions for open long-range integrable spin chains based on a given set of short-range integrable charges.
Motivated by the gauge/gravity duality, we study a model based on \cite{Beisert:2005wv,Beisert:2008cf} with local operators $\loc=\sum_a\loc(a)$ that act homogeneously on the spin chain sites $a$ (cf.\ \Figref{fig:locop}) and are invariant under a Lie (super)algebra $\alg{g}$. The symmetry $\alg{g}$ is assumed not to be broken by the boundary conditions. The derivations below were accompanied by computer verifications at the example of a $\alg{gl}(N)$ spin chain.

%

\section{Recursion for Periodic Chains }

Let us briefly review the recursive construction of periodic long-range spin chains \cite{Bargheer:2008jt,Bargheer:2009xy}:
The boost operator in \eqref{eq:srboost} is usually written as $\boost{\loc}=\sum_a a\loc(a)$. While it is not well-defined on periodic (but on infinite) spin chains, the commutator $\comm{\boost{\charge_2}}{\charge_r}$ is. Interestingly, the boost is merely a special case of a more general class of bilocal operators which are defined as compositions of two local operators $\loc_1$ and $\loc_2$, cf.\ \Figref{fig:biop}:
\begin{equation}
\biloc{\loc_1}{\loc_2}=\sum_{a+\range{\loc_1}<b}
\loc_1(a)\loc_2(b).
\label{eq:defbiloc}
\end{equation}
Here $\range{\loc}$ denotes the interaction range of $\loc$. In fact, the boost can be written as the composition of the identity $\id$ and a local operator in the form $\boost{\loc}=\biloc{\id}{\loc}$ such that the indentity counts the number of sites in front of $\loc$.
This generalization leads to a set of generators $\conn$ that may be applied to the Hamiltonian $\charge_2$ and the higher integrable charges $\charge_r$. These are local ($\conn_\mathrm{loc}=\loc$), boost ($\conn_\mathrm{boost}=\biloc{\id}{\charge_r}$) and bilocal ($\conn_\mathrm{bi}=\biloc{\charge_r}{\charge_s}$) generators which are distinguished from other candidates by the fact that they preserve locality and homogeneity. This locality is guaranteed since both legs of the bilocal generators individually commute with the charges $\charge_r$. 

We deform a set of short-range integrable charges defined for instance through \eqref{eq:srboost} by a differential equation:
\begin{equation}\label{eq:longdgl}
\frac{d}{d\xi}\charge_r(\xi)=i\comm{\conn(\xi)}{\charge_r(\xi)},
\qquad
r,s=2,3,\dots.
\end{equation}
If the initial condition to this equation is given by a short-range integrable system with charges obeying $\comm{\charge_r(0)}{\charge_s(0)}=0$, the solutions to \eqref{eq:longdgl} furnish a so-called perturbatively long-range integrable model \cite{Beisert:2003tq}:
\begin{equation}\label{eq:longcomm}
\comm{\charge_r(\xi)}{\charge_s(\xi)}=0,
\qquad 
r,s=2,3,\dots.
\end{equation}
Here the charges are perturbative series in the small parameters $\xi(\conn)$ associated with the generators $\conn$
\begin{equation}
\charge_r(\xi)=\sum_{k=0}^\infty\xi^k\charge_r^{(k)}.
\end{equation}
In fact, multiple deformation generators $\conn_\ell$ can be combined, resulting in charges depending on multiple parameters $\xi_\ell$.
In the context of the gauge/gravity duality the deformation parameters become specific functions of the coupling constant $\xi_\ell=\xi_\ell(\lambda)$. With higher orders of the parameters $\xi_\ell$ or the coupling $\lambda$, respectively, the interaction range of the charge operators typically increases. This implies that for a given chain, the interaction range necessarily exceeds the number of spin chain sites at a certain perturbative order. Beyond this so-called wrapping order, it is not known how to define a non-trivial long-range spin chain model of gauge/gravity type. Also this paper will not go beyond this asymptotic regime. 

\paragraph{Non-Triviality.}
Without specifying the operator $\conn$ in \eqref{eq:longdgl}, the induced deformations a priori look like similarity transformations with no impact on the spectrum. The key point is that $\conn$ can be well-defined on infinite but not on periodic chains, while $\comm{\conn}{\charge_r}$ is well-defined on both. This is the case for boost and bilocal charges, which require an ordering of the spin chain sites. 
Hence, equation \eqref{eq:longdgl} does not constitute a similarity transformation on the periodic chain if $\conn$ is chosen to be a bilocal operator.
Local operators $\conn$, well-defined on infinite and periodic chains, induce similarity transformations on both.
Details on the periodic construction as well as explicit expressions for the recursively generated long-range deformations of an $\alg{su}(2)$ symmetric Heisenberg spin chain are given in \cite{Bargheer:2008jt,Bargheer:2009xy}. This example corresponds to higher loop deformations of the dilatation generator in the $\alg{su}(2)$ sector of $\superN=4$ super Yang--Mills theory.

\section{Recursion for Open Chains}

Open boundary conditions only allow for half of the conserved charges $\charge_{2r}$. The odd charges defined on a periodic chain commute only up to boundary terms
\begin{equation}\label{eq:oddcomm}
\comm{\charge_{2r+1}}{\charge_s}=\loc^\bound_\Left+\loc^\bound_\Right,\quad r=1,2,\dots, s=2,3,\dots.
\end{equation}
The local boundary operators $\loc^\bound_{\Left/\Right}$ are non-vanishing only at the left or right boundary of the open chain, respectively, and vanish on a chain without boundaries. For the left side acting at site $a$ they take the form
\begin{equation}\label{eq:boundterms}
\loc^\bound_\Left(a)=\specleg\loc(a)-\loc(a),
\end{equation}
where $\specleg\loc(a):=\id(a)\otimes \loc(a+1)$ contains a spectator leg. Another important class of operators are spanning terms:
\begin{equation}\label{eq:FLterms}
\loc^\bound_{\Span}=\specleg\loc\specleg-\specleg\loc-\loc\specleg+\loc.
\end{equation}
They vanish on all spin chains except for those of length $\range{\loc}$ where they act on both boundaries at the same time. In the asymptotic regime considered here, the spin chain length is assumed to exceed the range of the operators. Hence, these contributions vanish.

If we want to define recursion relations for chains with open boundaries, two main puzzles arise: 
Firstly, bilocal operators with odd charges cannot induce local deformations in the above way, since the odd charges do not commute anymore.
Still the respective parameters appear in the open chain's perturbative spectrum \cite{Beisert:2008cf}. Secondly, the even charges acquire additional boundary terms on the open chain that cannot be generated on an infinite chain without boundaries.


In order to overcome these problems, we introduce the notion of semi-infinite spin chains. These are chains with an open boundary on one side of the chain and infinite extent on the other side. They will allow us to insert the odd charges on the bulk leg of a bilocal operator shielded from the boundary by the leg towards the boundary. In what follows we will mainly discuss \emph{left-open} chains, i.e.\ chains with an \emph{open boundary on the left}  and infinite extent on the right side. All quantities for the right-open case can be obtained using the parity transformation. 

\begin{figure}[t]
\Large $\sum\limits_{a+\range{\charge_{2r}}<b}$\quad\includegraphicsbox[scale=.8]{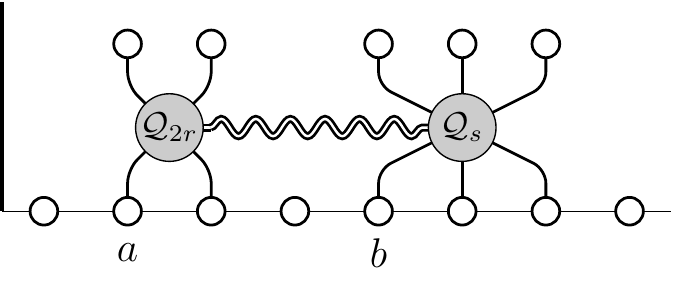}
\caption{Bilocal operator acting close to the left boundary.}
\label{fig:biop}
\end{figure}

The crucial idea is to generate charge deformations on the left- and right-open chain and to combine these deformations to operators commuting on a finite open chain. We therefore modify equation \eqref{eq:longdgl} according to
\begin{equation}\label{eq:longdglLR}
\frac{d}{d\xi}\charge_{r,\Left/\Right}(\xi)=i\comm{\conn_{\Left/\Right}(\xi)}{\charge_{r,\Left/\Right}(\xi)}|_{\Left/\Right},
\end{equation}
where $|_{\Left/\Right}$ denotes the application of open boundary conditions, i.e.\ elimination of terms like \eqref{eq:boundterms} for the left or right side, respectively.
Boost generators $\boostLeft{\charge_{2s+1}}=\biloc{\id}{\charge_{2s+1}}$ or $\boostRight{\charge_{2s+1}}=\biloc{\charge_{2s+1}}{\id}$ may now be used on the left- or right-open chain, respectively. By definition of the bilocal operator, the charge $\charge_{2s+1}$ will not see the boundary of the chain since it is shielded by the identity operator. Hence, the odd charges, which commute with the even charges in the bulk of the chain, may be used for deformation.
Similarly we can use the even charges inserted into a bilocal operator $\biloc{\charge_{2r}}{\charge_{2s+1}}$ or $\biloc{\charge_{2s+1}}{\charge_{2r}}$ as a buffer towards the boundaries, cf.\ \Figref{fig:biop}. 
In this construction the odd charges defined modulo boundary terms thus only live in the bulk of the chain.


We will now explain how to combine the charge structures generated on the semi-infinite chains in order to obtain commuting long-range charges on a finite open chain. Consider an arbitrary charge term on the left- or right-open spin chain composed of a bulk and boundary contribution and commuting on the respective chain:
\begin{equation}\label{eq:lrterms}
\charge_{2r,\Left/\Right}=\charge_{2r}^\bulk+\charge^\bound_{2r,\Left/\Right}.
\end{equation}
Here the bulk part is defined to contain no spectator legs and the boundary term vanishes in the bulk.
The following argument applies to all pairs of charge terms which have the same bulk structure $\charge_{2r}^\bulk$ in the left- and right-open case. For the left-open semi-infinite spin chain, we can expand the vanishing commutator 
\begin{equation}
\comm{\charge_{2r,\Left}}{\charge_{2s,\Left}} |_\Left=\loc^\bound_\Right |_\Left= 0
\end{equation} 
in terms of the summands in \eqref{eq:lrterms} to find 
\begin{align}
&\comm{\charge_{2r}^\bulk}{\charge_{2s}^\bulk}=\loc^\bound_\Left+\loc^\bound_\Right,
\\
&\comm{\charge_{2r}^\bulk}{\charge_{2s,\Left}^\bound}
+
\comm{\charge_{2r,\Left}^\bound}{\charge_{2s}^\bulk}
+
\comm{\charge_{2r,\Left}^\bound}{\charge_{2s,\Left}^\bound}=-\loc^\bound_\Left.\nonumber
\end{align}
Here $\loc^\bound_{\Left/\Right}$ are boundary terms acting on the left or right boundary only. 
In order to promote the charges \eqref{eq:lrterms} to an integrable model on a finite open chain, we define
\begin{equation}\label{eq:fincharge}
\charge_{2r}=\charge_{2r}^\bulk+\charge_{2r,\Left}^\bound+\charge_{2r,\Right}^\bound.
\end{equation}
This definition implies
\begin{equation}\label{eq:commfin}
\comm{\charge_{2r}}{\charge_{2s}}
=
\comm{\charge_{2r,\Left}^\bound}{\charge_{2s,\Right}^\bound}
+
\comm{\charge_{2r,\Right}^\bound}{\charge_{2s,\Left}^\bound}=\loc^\bound_{\Span}
\simeq0.
\end{equation}
The commutators on the right hand side of \eqref{eq:commfin} 
represent
spanning terms \eqref{eq:FLterms} and vanish in the regime of asymptotic spin chains discussed here. The so-defined charges $\charge_{2r}$ thus commute on the finite open chain. Equation \eqref{eq:commfin} explicitly demonstrates how the definition of long-range integrable spin chains breaks down at the order where the charge operators span the whole state.
This is a characteristic feature of open \emph{and} periodic gauge/gravity long-range spin chains.

As mentioned above, for these arguments to work it is important that the left- and right-open charge deformations in \eqref{eq:lrterms} have the same bulk structure. To guarantee this for all deformations, we modify the definition of the bilocal generator by adding a local contribution to \eqref{eq:defbiloc}:
\begin{equation}
\biloc{\loc_1}{\loc_2}=\dots+\ueber{\loc_1}{\loc_2}.
\end{equation}
Here the local overlap of $\loc_1$ and $\loc_2$ is defined as
\footnote{This regularization was also chosen to render the interaction range of \emph{periodic} long-range charges minimal \cite{Bargheer:2009xy}.}
\begin{equation}\label{eq:bireg}
\ueber{\loc_1}{\loc_2}=\sum_{b\geq a-\ell_{12}}^{a+\range{\loc_1}}
(1-\half\delta_{a-b,\ell_{12}})
\half\{\loc_1(a),\loc_2(b)\},
\end{equation}
with $\ell_{12}=\half(\range{\loc_2}-\range{\loc_1})$.
The regularized operators obey
\begin{equation}
\biloc{\charge_r}{\charge_s}+\biloc{\charge_s}{\charge_r}=\charge_r\charge_s,
\end{equation}
which implies that $\biloc{\charge_r}{\charge_s}$ and $-\biloc{\charge_s}{\charge_r}$ induce the same bulk structure on the left- and right-open chain. The same applies to the boost charges, cf.\ \Tabref{tab:lrgens}.
Furthermore the dispersion relation of $\ueber{\charge_r}{\charge_s}$ equals the dispersion relation of $\biloc{\charge_r}{\charge_s}$. Local odd charges $\charge_{2r+1}$ generate pure boundary terms \eqref{eq:oddcomm}, i.e.\ the bulk part on the left- and right-open chain is trivially equal. For the local charges the relative minus sign in \Tabref{tab:lrgens} implies the non-triviality of the deformation. A plus results in a similarity transformation. 

Let us give a brief example for illustration: The integrable charges of an $\alg{su}(2)$ symmetric Heisenberg spin chain with open boundaries can be written in terms of permutation symbols. In particular, the Hamiltonian takes the form
\footnote{We do not identify the operator consisting of two spectator legs according to $\eqref{eq:boundterms}$ since it is useful to give the charges a vanishing vacuum eigenvalue.}
\begin{equation}
\charge_2^{(0)}=
\twospecleg-\permsym=\sum_{k=1}^{L-1}\big(\twospecleg_k-\permsym_k\big).
\end{equation}
Now we choose one of the generators $\conn$ in \Tabref{tab:lrgens}, e.g.\ the first odd charge $\charge_3^{(0)}=\ihalf (\threeonetwo -\twothreeone)$ of the periodic Heisenberg chain, and plug it into \eqref{eq:longdglLR}. This yields deformations for the left- and right-open case, e.g.\ at first order
\begin{equation}
 \charge_{2,\Left}^{(\delta_3)}=i\comm{\charge_3^{(0)}}{\charge_2^{(0)}}|_\Left= \permsym-\onethreetwo.
\end{equation}
Adding the left- and right-open contributions according to \eqref{eq:fincharge}, we thus find 
\begin{equation}
\charge_2(\delta_3)=\twospecleg-\permsym+\delta_3\bigbrk{2\permsym-\onethreetwo-\twoonethree}+\order{\delta_3^2}.
\end{equation}
Similarly we can deform all short-range charges $\charge_{2r}^{(0)}$ using any generator $\conn$ in \Tabref{tab:lrgens}. The resulting charges perturbatively commute on the finite open spin chain up to wrapping order.

\begin{table}[t]
\begin{tabular}{|l|c|c|c|}\hline
Generator $\conn$&Left-Open&Right-Open&Parameter\\\hline
Boost Charge&$\biloc{\id}{\charge_{2r+1}}$& $-\biloc{\charge_{2r+1}}{\id}$&$\alpha$\\\hline
Bilocal Charge&$\biloc{\charge_{2r}}{\charge_{2s+1}}$&$-\biloc{\charge_{2s+1}}{\charge_{2r}}$&$\beta$\\\hline
Basis Change&$\rot_{r,s}$&$\rot_{r,s}$&$\gamma$\\\hline
Local Charge&$\charge_{2r+1}$&$-\charge_{2r+1}$&$\delta$\\\hline
\end{tabular}
\caption{Generators on the left- and right-open spin chain.}
\label{tab:lrgens}
\end{table}

\paragraph{Non-Triviality.}
Bilocal operators require an ordering of the spin chain sites. Consequently, they are in general \emph{not} compatible with \emph{periodic} boundaries and this incompatibility rendered the periodic deformations induced by \eqref{eq:longdgl} non-trivial. On the other hand, bilocal operators \emph{are} compatible with \emph{open} boundary conditions. Why are the deformations defined by \eqref{eq:longdglLR} still non-trivial? In the open case the non-triviality stems from the fact that we deform the charges by adding two contributions on which \emph{either} left- or right-open boundary conditions were applied, cf.\ \eqref{eq:fincharge}. In general, a corresponding transformation cannot be performed on a finite open chain which has boundaries on both sides. It is thus non-trivial. The construction with left- and right-open spin chains is necessary for deformation generators including the odd charges. Deformations induced by even charges can be equally well performed on a finite open chain. Thus, even local ($\charge_{2r}$), boost ($\boost{\charge_{2r}}$) or bilocal ($\biloc{\charge_{2r}}{\charge_{2s}}$) charges indeed correspond to similarity transformations
\footnote{In \cite{Beisert:2008cf} the parameters corresponding to even boosts were not yet identified as similarity transformations due to taking different linear combinations of the charges.}.

\section{Deformations \& Bethe Ansatz}
We now derive the open long-range Bethe ansatz. We discuss on the left-open chain how the different deformations modify one- and two-magnon eigenstates of the charges and their one-magnon eigenvalues $q_r(p)$. Here we consider an $\alg{su}(2)$ spin chain with spin $t/2$ representation on all sites. The generalization to higher rank algebras follows in analogy to the periodic case given in \cite{Bargheer:2008jt,Bargheer:2009xy}. 

\emph{Boost operators $(\alpha)$} of odd charges serve as generators of long-range deformations 
via the equation (cf.\ \eqref{eq:longdglLR})
\begin{equation}\label{eq:boostdefexp}
\frac{d}{d\alpha_{2k+1}}\charge_{r}=i \comm{\boostLeft{\charge_{2k+1}}}{\charge_{r}}|_\Left.
\end{equation}
If we evaluate a boosted charge on a one-magnon state $\ket{p}$ in the bulk of the spin chain \cite{Bargheer:2009xy}
\begin{equation} 
\bigcomm{\boostLeft{\charge_{2k+1}}}{\charge_r}\ket{p}=iq_{2k+1}(p)\frac{\partial q_r(p)}{\partial p}\ket{p},
\end{equation}
we find that boost deformations \eqref{eq:boostdefexp} imply a differential equation for the one-magnon charge eigenvalues
\begin{equation}\label{eq:dglqboost}
\frac{dq_r(p)}{d\alpha_{2k+1}} 
=- q_{2k+1}(p)\frac{\partial q_r(p)}{\partial p} .
\end{equation}
Introducing an integration constant $t$ typically labeling the spin representation, this equation is solved by 
\begin{equation}\label{eq:chargeev}
q_r(u,t)=\frac{i}{r-1}\left(\frac{1}{(u+\ihalf t)^{r-1}}-\frac{1}{(u-\ihalf t)^{r-1}}\right).
\end{equation}
Here we implicitly define the rapidity $u$ and the rapidity map $x(u)$ associated with the momentum $p$ by
\footnote{This definition of $x(u)$ differs from the one that renders the interaction range of periodic long-range charges minimal \cite{Bargheer:2009xy}. Here $dq_r(u)/d\alpha_k=0$ since $\gamma=0$.}
\begin{equation}\label{eq:defalpha}
e^{ip(t,u)}=\frac{x(u+\ihalf t)}{x(u-\ihalf t)},
\quad x(u)=u\exp\bigg(-\sum_{k=1}^\infty\frac{\alpha_{2k+1}}{2ku^{2k}}\bigg).
\end{equation}

\emph{Bilocal deformations $(\beta)$} are generated by 
\begin{equation}\label{eq:bideform}
\frac{d}{d\beta_{2r,2s+1}}\charge_t=i\comm{\biloc{\charge_{2r}}{\charge_{2s+1}}}{\charge_t}|_\Left.
\end{equation}
In the bulk, the action of a bilocal charge on an ordered two-magnon state is given by $\biloc{\charge_r}{\charge_s}\ket{u<u'}=q_r(u)q_s(u')\ket{u<u'}$, where we neglect local contributions \eqref{eq:defbiloc} whose impact is discussed below. To build an asymptotic two-particle eigenstate, we define
\begin{equation}\label{eq:twoeigen}
\ket{u,u'}= \ket{u'<u}+S(u,u')\ket{u<u'}.
\end{equation}
Then \eqref{eq:bideform} induces a differential equation on $S(u,u')$ \cite{Bargheer:2009xy}
\begin{equation}
\frac{dS(u,u')}{d\beta_{2r,2s+1}}\ket{u<u'}=i\biloc{\charge_{2r}}{\charge_{2s+1}}\ket{u<u'}
\end{equation}
which is solved by 
the two-particle scattering factor
\begin{equation}
S(u,u')=e^{-2i\theta_{2r,2s+1}(u,u')}S_{\SR}(u-u').
\end{equation}
Here $S_{\SR}(u-u')=(u-u'-i)/(u-u'+i)$ denotes the undeformed scattering factor and the so-called dressing phase is given by \cite{Arutyunov:2004vx,Beisert:2005wv}
\begin{equation}\label{eq:dressphas}
\theta_{r,s}(u,u')=\beta_{r,s}\bigbrk{q_{r}(u)q_{s}(u')-q_{s}(u)q_{r}(u')}.
\end{equation}

\emph{Basis changes $(\gamma)$} of the charges are important to adjust the interaction range of higher order charge terms \cite{Bargheer:2009xy}. 
They are implemented by introduction of an associated parameter class $\gamma$ and a rotation generator $\rot_{r,s}$:
\begin{equation}
\frac{d}{d\gamma_{r,s}}\charge_{r}=\comm{\rot_{s,r}}{\charge_{r}},
\end{equation}
with $\comm{\rot_{s,r}}{\charge_r}=\charge_s$ and $\gamma_{\mathrm{even},\mathrm{odd}}=0$. We do not discuss the range of higher order deformations and assume $\gamma=0$.

\emph{Local Operators $(\delta)$} can deform the spectrum of open as opposed to periodic chains non-trivially. When a magnon is reflected at a boundary, its momentum or rapidity flips sign, i.e.\ $u\to \bar u=-u$ 
\footnote{If similarity transformations with the even boosts are switched on, the reflection map $p\to -p$ is deformed by the corresponding parity breaking parameters, cf.\ \cite{Beisert:2008cf}.}.
 In analogy to \eqref{eq:twoeigen} we define an asymptotic boundary scattering state
$\ket{u,\bar u}= \ket{\bar u}+S_\Left(u)\ket{u}$
and rephrase the statement that local operators $\loc_{A,\Left}$ induce a phase in form of a differential equation
\begin{equation}\label{eq:bounddgl}
\frac{d S_\Left(u)}{d\delta_{A,\Left}}\ket{u}=i\loc_{A,\Left}\ket{u},
\end{equation}
where in the bulk $\loc_{A,\Left}\ket{u}=\ell_{A,\Left}(u)\ket{u}$. Equation \eqref{eq:bounddgl} is solved by the boundary scattering factor
\begin{equation}
S_{\Left}(u)=e^{2i\phi_\Left(u)}S_{\SR,\Left}(u),
\end{equation}
with $\phi_\Left(u)=\delta_{A,\Left}(\ell_{A,\Left}(u)-\ell_{A,\Left}(\bar u))$. The only local operators $\loc_{A,\Left}$ contributing to the integrable deformations above are the odd charges $\charge_{2t+1}$, i.e.\ the regulator \eqref{eq:bireg} of the boosts, as well as via \eqref{eq:bideform} the local regulator of the bilocal charges $\ueber{\charge_{2r}}{\charge_{2s+1}}$.
Using $q_{2t+1}(\bar u)=-q_{2t+1}(u)$, the boundary phase takes the form
\begin{equation}
\phi_{\Left}(u)= \delta_{2t+1,\Left}q_{2t+1}(u)-\theta_{2r,2s+1}(u,\bar u),
\end{equation}
including a boundary part of the dressing phase \eqref{eq:dressphas}.

The \emph{Bethe equations} have to be satisfied by roots $u_{k}$, $k=1,\dots,M$ describing eigenstates of the finite chain:
\begin{align}
&e^{i\brk{p_{k}-\bar p_{k}}\,L}
=S_\Left(u_{k})S_\Right(\bar u_{k})\mathop{\prod_{j=1}^{M}}_{j\neq k}\nonumber
S(u_{k},u_{j})S^{-1}(\bar u_{k},u_{j}).
\label{eq:openbethegen}
\end{align}
Above, we have derived the corresponding deformations of the bulk and boundary scattering factors according to 
\begin{align}
 &S(u,u')S^{-1}(\bar u,u')
 = S_\SR(u,u')S_\SR^{-1}(\bar u,u') e^{2 i( \theta(u,u')-\theta(\bar u,u'))},\nonumber\\
&S_\Left(u)S_\Right(\bar u)
=S_{\Left,\SR}(u)S_{\Right,\SR}(\bar u)e^{2i\brk{\phi_\Left(u)-\phi_\Right(\bar u)}}.
\end{align}
Due to integrability multi-magnon eigenvalues are given by sums over one-magnon eigenvalues.
Importantly, these Bethe equations for the finite open spin chain are merely asymptotic, i.e.\ valid for chains longer than the range of the charges. 
\section{Conclusions}
Here we have shown how to generate long-range integrable charge operators 
as deformations of short-range models defined on open spin chains.
All deformations presented above can be combined yielding expansions in multiple moduli, e.g.\ 
$\charge_{2r}=\charge_{2r}^{(0)}+\alpha_3\beta_{2,3}\charge_{2r}^{(3,2|3)}+\dots.$
Different choices of the parameters $\alpha$, $\beta$, $\gamma$ and $\delta$ correspond to different integrable systems covering the whole moduli space found in \cite{Beisert:2008cf}. The existence of this long-range recursion is remarkable since so far neither for short- nor for long-range open spin chains an iterative construction was known. 
While the long-range formalism for periodic and open chains is based on \eqref{eq:longdgl} or \eqref{eq:longdglLR}, respectively, the periodic short-range recursion \eqref{eq:srboost} relies on the transfer matrix formalism which significantly differs for the open chain \cite{Grabowski:1996qf}. 
It would be highly desirable to find a way of extending the presented open long-range recursion to the short-range case. 
This could eventually unveil how long-range integrable spin chains fit into the framework of standard integrable models.
A study of the interaction range of the charges at higher orders including a detailed investigation of the flatness of moduli space would be important to reproduce the range patterns found in \cite{Beisert:2008cf}.

\paragraph{Acknowledgements.}
The author is very grateful to Till Bargheer and Niklas Beisert for previous work as well as for many helpul discussions and suggestions.

\bibliography{openboost}

\end{document}